\documentclass[aip,reprint]{revtex4-1}

\usepackage{graphicx}
\usepackage[utf8]{inputenc}
\usepackage[T1]{fontenc}
\usepackage{mathptmx}
\usepackage{etoolbox}

\draft 
\newcommand{\alscn}{Al\textsubscript{0.92}Sc\textsubscript{0.08}N~}
\newcommand{\alscnNospace}{Al\textsubscript{0.92}Sc\textsubscript{0.08}N}
\newcommand{\alscnx}{Al\textsubscript{1-x}Sc\textsubscript{x}N~}
\newcommand{\alscnxNospace}{Al\textsubscript{1-x}Sc\textsubscript{x}N}
\newcommand{\alscnM}{Al\textsubscript{0.85}Sc\textsubscript{0.15}N~}
\newcommand{\alscnMNospace}
{Al\textsubscript{0.85}Sc\textsubscript{0.15}N}

\begin{document}

\preprint{}

\title{Interfacial Polarization Switching in \alscnNospace/GaN Heterostructures Grown by Sputter Epitaxy}

\makeatletter
\def\@email#1#2{
 \endgroup
 \patchcmd{\titleblock@produce}
  {\frontmatter@RRAPformat}
  {\frontmatter@RRAPformat{\produce@RRAP{*#1\href{mailto:#2}{#2}}}\frontmatter@RRAPformat}
  {}{}
}
\makeatother


\author{Niklas Wolff*}
\email[]{niwo@tf.uni-kiel.de}
\affiliation{Department of Material Science, Kiel University, Kaiserstr. 2, D-24143 Kiel, Germany}
\affiliation{Kiel Nano, Surface and Interface Science (KiNSIS), Kiel University, Christian-Albrechts-Platz 4, D-24118 Kiel, Germany}

\author{Georg Schönweger}
\affiliation{Department of Material Science, Kiel University, Kaiserstr. 2, D-24143 Kiel, Germany}
\affiliation{Fraunhofer Institute for Silicon Technology (ISIT), Fraunhoferstr. 1, D-25524 Itzehoe, Germany}

\author{Redwanul Md. Islam}
\affiliation{Department of Material Science, Kiel University, Kaiserstr. 2, D-24143 Kiel, Germany}

\author{Ziming Ding}
\affiliation{Advanced Electron Microscopy in Materials Research, Institute of Nanotechnology (INT), Karlsruhe Institute of Technology (KIT), D-76344 Eggenstein-Leopoldshafen, Germany}
\affiliation{Karlsruhe Nano Micro Facility (KNMFi), Karlsruhe Institute of Technology (KIT), D-76344 Eggenstein-Leopoldshafen, Germany}

\author{Christian Kübel}
\affiliation{Advanced Electron Microscopy in Materials Research, Institute of Nanotechnology (INT), Karlsruhe Institute of Technology (KIT), D-76344 Eggenstein-Leopoldshafen, Germany}
\affiliation{Karlsruhe Nano Micro Facility (KNMFi), Karlsruhe Institute of Technology (KIT), D-76344 Eggenstein-Leopoldshafen, Germany}

\author{Simon Fichtner}
\affiliation{Department of Material Science, Kiel University, Kaiserstr. 2, D-24143 Kiel, Germany}
\affiliation{Kiel Nano, Surface and Interface Science (KiNSIS), Kiel University, Christian-Albrechts-Platz 4, D-24118 Kiel, Germany}
\affiliation{Fraunhofer Institute for Silicon Technology (ISIT), Fraunhoferstr. 1, D-25524 Itzehoe, Germany}

\author{Lorenz Kienle}
\affiliation{Department of Material Science, Kiel University, Kaiserstr. 2, D-24143 Kiel, Germany}
\affiliation{Kiel Nano, Surface and Interface Science (KiNSIS), Kiel University, Christian-Albrechts-Platz 4, D-24118 Kiel, Germany}


\date{\today}

\begin{abstract}
The integration of ferroelectric nitride thin films such as \alscnx onto GaN templates could enable enhanced functionality in novel high-power transistors and memory devices. This requires a detailed understanding of the ferroelectric domain structures and their impact on the electric properties. In this contribution, the sputter epitaxy of highly coherent \alscn thin films grown on GaN approaching lattice-matching conditions is demonstrated. Scanning transmission electron microscopy investigations reveal the formation of polar domains and the mechanism of domain propagation upon ferroelectric switching. Atomic resolution imaging suggests that polarization inversion is initiated by an interfacial switching process in which already the first atomic layer of \alscnx changes its polarization from the as-grown M- to N-polarity. An atomically sharp planar polarization discontinuity is identified at the \alscnNospace/GaN interface and described by atomic modeling and chemical structure analysis using electron energy loss spectroscopy, considering local lattice spacings. Moreover, residual domains with M-polarity are identified at the top Pt electrode interface. These insights on the location and the atomic structure of ferroelectric inversion domains in sputter deposited \alscnxNospace/GaN heterostructures will support the development of future non-volatile memory devices and novel HEMT structures based on ferroelectric nitride thin films via interface engineering. 

\end{abstract}

\maketitle 

\section{Introduction}\label{sec1}

The electronic properties resulting from the direction and magnitude of spontaneous polarization within multilayer structures of non-centrosymmetric crystal structures are widely exploited in the field of epitaxial group III-N heterojunctions. These heterojunctions are designed for optoelectronic applications or power electronic devices because of their exceptionally high electron or hole gas confinement and high mobilities created at these interfaces.\cite{Bastard1986, Kai2023, Holonyak1980} 

The formation of a two-dimensional electron gas (2DEG) between two semiconductors was initially achieved through the junction of GaAs and AlGaAs,\cite{Dingle1978, Störmer1979} pioneering the development of contemporary AlGaN/GaN-based high electron mobility transistor structures (HEMTs) for high-power and high-frequency applications.\cite{Kahn1993}
The design of HEMTs necessitates the optimization of interface properties. This is due to the fact that crystalline defects, including dislocations and lattice strain, have the potential to be detrimental to the properties of the 2DEG.\cite{Hori2013, Zhang1999, Choi2013, Gonschorek2006, Joshi2003, Zhang2000}

In the recent past, significant advancements in the field of thin film ferroelectrics have led to the discovery of novel materials, including doped hafnium oxide\cite{Böschke2011}, Sc- and B-substituted AlN\cite{Fichtner2019, Hayden2021} and Sc-substituted GaN\cite{Wang2021, Uehara2021}. The integration of ferroelectric functionality into semiconductor heterostructures is intriguing for the manipulation of interfacial sheet charges.\cite{Stern1967} This can be achieved by dynamically modulating the direction of spontaneous polarization in a ferroelectric layer positioned adjacent to an AlGaN/GaN heterostructure\cite{Li2022, Wu2020, Zhao2023, Wang2023, Okuda2025} or within the ferroelectric/GaN heterostructure itself.\cite{Casamento2022, Casamento2023} 

The new wurtzite-type ferroelectric nitrides, such as \alscnxNospace, hold particular promise for facilitating the development of all-nitride-based ferroelectric (FE)-HEMTs, featuring non-volatile memory functionality and/or normally-off operation. Theoretical calculations have demonstrated that polarization inversion can enhance electron sheet densities along \alscnxNospace/GaN interfaces by more than one order of magnitude, particularly in the presence of sharp domain walls with head-to-head configuration at the interface.\cite{Yassine2024}
However, the existence and exploitation of such domain walls in these systems remain to be elucidated.
Furthermore, the optimization of growth conditions and material parameters is imperative to achieve defect-poor, high crystal quality and electrically tunable interfaces. Numerous studies have focused on lattice-matching the \alscnx layer and the GaN-based template. However, the lattice-matched Sc composition of \alscnx on GaN has been reported to range from 9\% to 18\%.\cite{Deurzen2023, Casamento2020, Kumar2024, Dinh2023, Motoki2024, Nguyen2024, Wolff2024} This presents a substantial challenge, given that identifying the appropriate lattice-matched composition and its sensitivity to growth methods and conditions is imperative for the successful fabrication of high-quality, strain-free GaN-based heterostructures incorporating ultra-wide band-gap (ferroelectric) materials. 
In conjunction with the advancement of enhanced growth conditions, it is necessary to gain a profound comprehension of the atomic configuration within the integrated ferroelectric nitride layer, particularly concerning crystalline or chemical imperfections and residual lattice strain. Moreover, a comprehensive understanding of ferroelectric polarization discontinuities in III-N heterostructures is essential for the progression from fundamental research to device development. In this regard, the identification of the polarity at the unit cell level, the geometry of the polar domains, their propagation mechanisms, and the resulting local atomic structures at the domain walls before and after ferroelectric switching are pivotal parameters to be determined. This is particularly true for the polarization discontinuity at the interface, which is defining the electronic properties of the heterojunction.

With this contribution, we report on direct polarization investigations of ferroelectric nearly lattice-matched \alscnNospace/GaN heterostructures grown by physical vapor deposition (PVD - sputter epitaxy) featuring high structural coherency. High-resolution scanning transmission electron microscopy (HRSTEM) was performed to investigate the electric field-induced domain structures after \textit{ex situ} polarization inversion. We identify the stabilization of small M-polar nucleation centers at the Pt/\alscn interface subsequent to the conversion of the spontaneous polarization to N-polarity. Concurrently, a planar inversion domain boundary is formed at the GaN interface, marking the initial phase of the polarization inversion process. A comparison is made between these results and the domain structures and switching mechanisms observed in \alscnM layers that have been grown by the metal organic chemical vapor deposition (MOCVD) method.\cite{Wolff2025} The direct observation of domain boundary structures in sputter deposited \alscnx thin films will have a significant impact on the research and development of novel devices in the field of non-volatile memories and power electronics based on the ferroelectric switching properties of \alscnx thin films.

\section{Results and Discussions}\label{sec2}

\subsection{Nanostructure of epitaxial \alscnxNospace/GaN heterostructures}

\begin{figure*}[h!]
\centerline{\includegraphics[width=\textwidth]{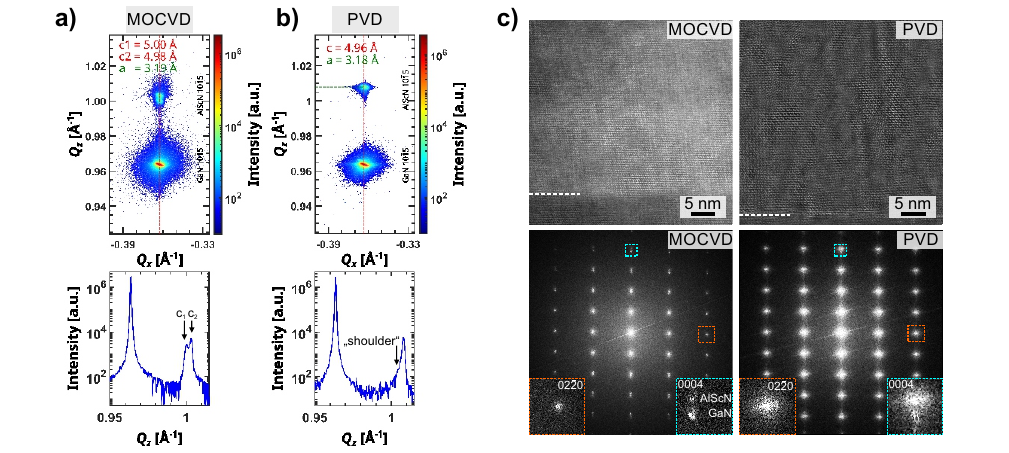}}
\caption{Comparison of nanostructure properties of ferroelectric and pseudomorphic \alscnxNospace/GaN thin film heterostructures produced by MOCVD (\alscnM, 230 nm) and PVD (\alscn, 110 nm). High-resolution RSM ($Q_x - Q_z$) of the AlScN and GaN 10\={1}5 reflections for a) MOCVD and b) PVD grown heterostructures. The intensity profiles show cutlines along $Q_z$ at the center position $Q_x$ of the reflections. c) HRTEM/FFT investigation of the crystalline properties at the GaN interface.\label{fig:nanostructure}}
\end{figure*}

The crystal quality and epitaxy in the pseudomorphic (i.e. growing with same in-plane lattice parameter) \alscnxNospace/GaN heterostrucutures grown by MOCVD and PVD is compared using high-resolution reciprocal space maps (RSM) of the asymmetric \alscnx and GaN 10\={1}5 reflections. These reflections have out-of-plane and in-plane components which make them ideal candidates to determine the \textit{c} and \textit{a} lattice parameters of the wurtzite-type structure, the latter giving access to compositional variation to achieve in-plane lattice matched conditions. The average composition of the MOCVD film was previously determined to be \alscnM by secondary ion mass spectroscopy and energy-dispersive X-ray spectroscopy (STEM-EDS).\cite{Wolff2024} By the evaluation of a series of sputter deposited films within the compositional range $x \sim$ 0.04 - 0.18, we found that films with a composition of $x \sim$ 0.08 and $\sim$ 0.10 determined by SEM-EDS show the best structural quality with comparable narrow $\omega$(0002) rocking curve full-width at half maximum of $\sim$ 258 and $\sim$ 261 arcsec, compared to 252 arcsec for the MOCVD grown layer. This is in agreement with some of the Sc compositions reported for lattice matched conditions in literature.\cite{Nguyen2024} 
The Sc content of a thin cross-section TEM lamella was determined to $x \sim$ 0.07 using STEM-EDS which is matching with $x \sim$ 0.08 determined using SEM-EDS within the anticipated error margin of $\sim$ 2 at.\%.
The RSMs for the MOCVD \alscnM and PVD \alscn grown heterostructures are displayed in Figure \ref{fig:nanostructure}a and \ref{fig:nanostructure}b and show great vertical alignment of the reflections on the $Q_x$-axis indicating in-plane lattice matching for both systems. However, a small deviation in the \textit{a} lattice parameter for \alscn with 3.18 $\mathring{A}$ from 3.19 $\mathring{A}$ (GaN) is determined. The lower Sc content in the PVD film also gives rise to a smaller \textit{c} lattice parameter of 4.96 $\mathring{A}$ compared to the MOCVD grown layer with higher Sc content. A small asymmetry of the \alscn 10\={1}5 reflection intensity is observed in the $Q_x$-intensity profile. The shoulder at smaller $Q_x$ values indicates the presence of areas with small tensile strain along the \textit{c}-axis, possibly at the GaN interface. Furthermore, a peculiarity which is a split of the 10\={1}5 reflection is observed for the MOCVD grown layer with \textit{$c_1$} = 5.00 $\mathring{A}$ and \textit{$c_2$} = 4.98 $\mathring{A}$. This observation might indicate a small compositional heterogeneity within the measured sample volume, which was not evidenced in our former TEM studies on a locally extracted lamella with dimensions of 10 $\mu$m \textit{x} < 100 nm, suggesting that such possible compositional domains may exist on larger length scales.\cite{Wolff2024} 
The crystal quality of the \alscnxNospace/GaN interface for the as-grown heterostructures is investigated by high-resolution TEM (HRTEM). The Fourier-filtered HRTEM micrographs recorded at an instrumental magnification of 300kX and their Fast-Fourier-Transforms (FFT) are presented in Figure \ref{fig:nanostructure}c. Although the larger-scale structural analysis via XRD show minimal differences in crystal quality between both thin films, the nanostructure analyses at the epitaxial interfaces show superior structural quality of the MOCVD-grown layer. This is apparent from the FFT patterns showing a very clear separation and a defined circular shape of the 0004 and 02\={2}0 reflection spots of the \alscnM layer and the GaN template in comparison to the elliptically shaped and more diffuse intensities in case of the sputtered \alscn layer. The diffuse broadening of the reflections is related to higher structural disorder apparent from the HRTEM micrographs. Nonetheless, the growth of a strongly \textit{c}-axis textured \alscn thin film with high structural coherence is achieved using sputter epitaxy at close to lattice-matched film compositions.

\begin{figure*}[h!]
\centerline{\includegraphics[width=\textwidth]{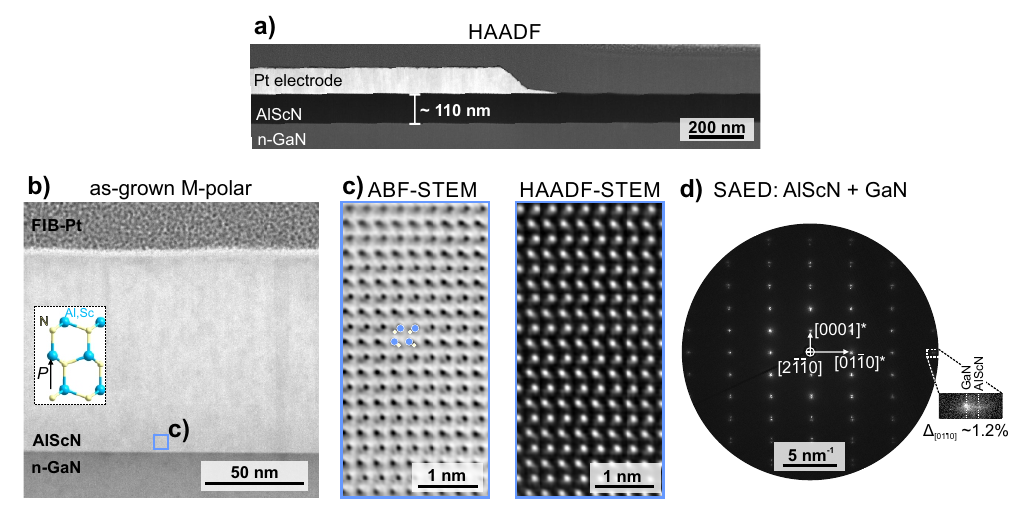}}
\caption{TEM investigation of the as-grown thin film heterostructure. a) HAADF-STEM image of the Pt/\alscnNospace/n-GaN capacitor in cross-section. b) ABF-STEM image of the as-grown M-polar \alscnNospace/GaN region. The sketch shows the M-polar atomic structure for the [2\={1}\={1}0] crystal orientation. c) Atomic-resolution HRSTEM images showing M-polar orientation near the GaN interface. d) SAED pattern in [2\={1}\={1}0] zone axis orientation.\label{fig:Mpolar}}
\end{figure*}

Next, we focus on the nanostructure and atomic scale investigations of the PVD \alscn layer to determine the as-grown film polarity using STEM. A cross-section high-angle annular dark-field (HAADF)-STEM image of the Pt/\alscnNospace/n-GaN capacitor structure is displayed in Figure \ref{fig:Mpolar}a showing a layer thickness of 110 nm and a smooth surface after structuring of the Pt electrode by ion-beam etching. The edge profile of the electrode does not end abruptly but is tapered in extending for about 100 nm after the end of the diagonal facet. Figure \ref{fig:Mpolar}b shows the annular bright-field (ABF)-STEM micrograph of the as-grown part of the cross-section lamella. On the atomic scale, the film appears highly coherent, which enabled the clear determination of metal (M)-polarity from high-resolution (HRSTEM) micrographs recorded from the top interface, the film bulk and the GaN interface (cf. Figure \ref{fig:Mpolar}c). The identified unipolar structure is different compared to layers with higher Sc content, where a flip to the N-polar structure was observed after 40\,-\,50 nm in the as-grown state, presumably by strain relaxation.\cite{Schönweger2022} The N-atomic column positions were as clearly imaged with the HAADF-STEM signal but less influenced by the nanoscale inhomogeneities within the crystal lattice, which is leading to the smeared reflections as visible in Figure \ref{fig:nanostructure}c. Thus, this imaging mode is dominantly used for studies of the local polar domain structure conducted on ferroelectrically switched capacitors (see Section \ref{subsec.domains}.)

In analogy to the XRD examinations, a selected-area electron diffraction (SAED) experiment was performed on the \alscnNospace/GaN heterostructure with the investigated region limited by the diffraction aperture having a virtual circular diameter of 250 nm. Figure \ref{fig:Mpolar}d shows the single crystalline reflection patterns of the GaN and \alscn film in [2\={1}\={1}0] zone axis. The observed splitting of the out-of-plane reflections along [0001]* agrees with the different \textit{c}-axis lattice parameters of GaN and \alscnNospace. However, by inspection of the splitting along the in-plane reflections a difference of about -1.2\% is evidenced in between the 01\={1}0 reflections, which indicates non-perfect in-plane lattice-matched epitaxial growth.

\subsection{Ferroelectric domains in switched N-polar \alscn thin films}\label{subsec.domains}

\begin{figure*}[h!]
\centerline{\includegraphics[width=15cm]{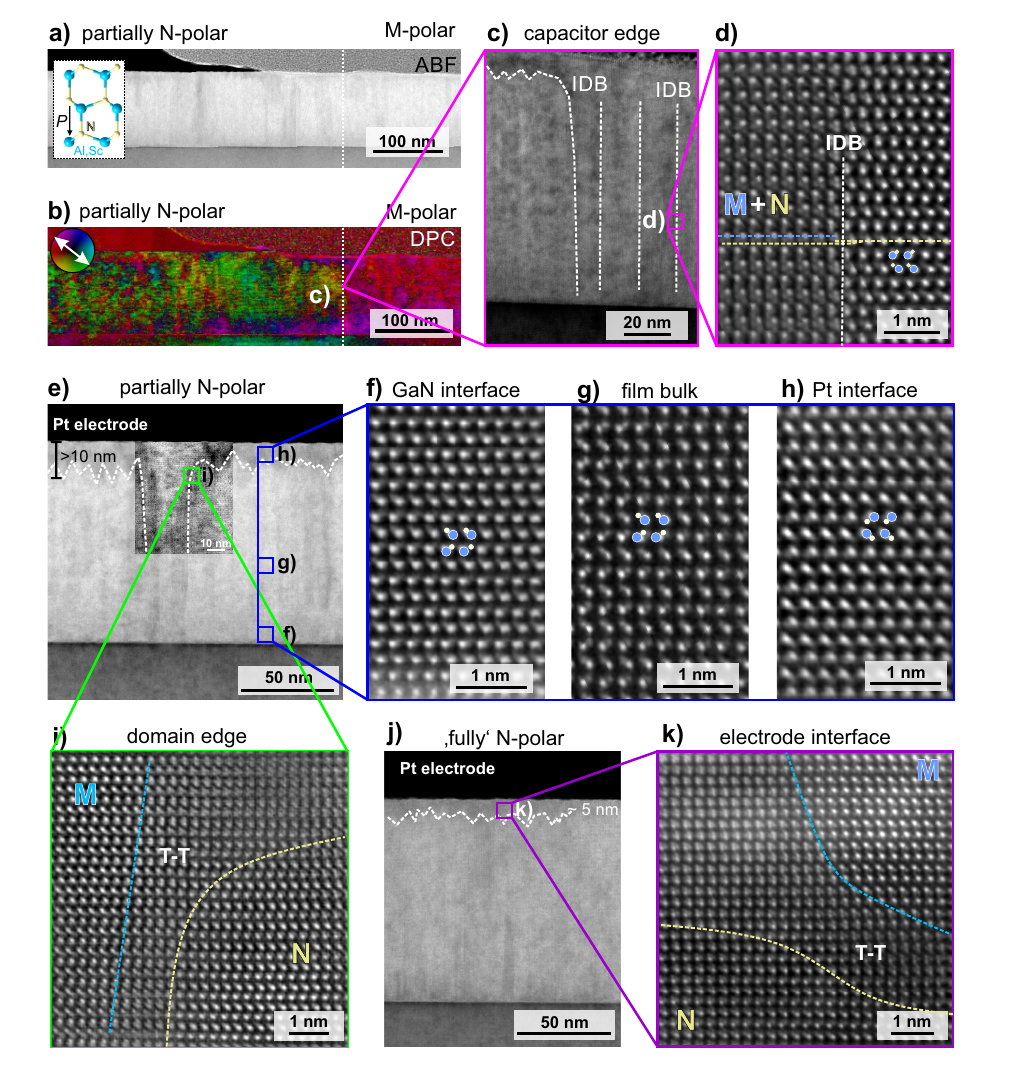}}
\caption{Investigation of ferroelectric domain structures after partial $E_{switch}$ < $E_c$ and complete $E_{switch}$ > $E_c$  polarization inversion. a) ABF-STEM cross-section image and b) DPC image of a sub-saturation field biased Pt/\alscnNospace/n-GaN capacitor with the partially switched N-polar layer (green: left) and the as-grown M-polar film (red: right). c) ABF-STEM image showing the edge-region of the switched capacitor with multiple inversion domain boundaries. d) HRSTEM image showing a vertical inversion domain boundaries at the position where a DPC contrast change was observed. e) ABF-STEM micrograph showing the polar domain structure after partial switching The dashed lines indicate domain boundaries. f-h) Investigation of local polarity by ADF-HRSTEM at the GaN interface, the film bulk and the Pt interface. i) HRSTEM image of a curved tail-to-tail domain wall. j) ABF-STEM image of a fully switched N-polar layer. k) HRSTEM micrograph showing nanoscale residual M-polar domains stabilized at the Pt-interface after complete switching.\label{fig:Npolar}}
\end{figure*}

The ferroelectric domain structure in nearly lattice-matched wurtzite-type \alscnx thin films can be investigated on \textit{ex situ} switched layers by a combination of scattering contrast change in ABF-STEM and validation of domain boundaries by HRSTEM.\cite{Wolff2024, Wolff2025} By comparing as-grown films with films to which non-saturating and saturating voltage signals were applied to partially and completely invert the polarization direction, the domain propagation and hence the switching mechanisms can be inferred. The ferroelectric pretreatment is presented in Figure S1 (Supporting Information). In Figure \ref{fig:Npolar} we provide an in-depth analysis of partially-switched and fully-switched films after the application of a positive voltage signal to switch from the prior set M-polar to the N-polar state after cycling through 20 complete hysteresis loops. The ABF-STEM micrograph presented in Figure \ref{fig:Npolar}a shows the partially switched layer within the capacitor structure in comparison to the as-grown layer on the right. The ABF-STEM micrograph contains contrast changes arising from local alterations of diffraction conditions, for example by minute tilts of crystal orientation or enhanced scattering by a change of atomic density present at the overlap of  M-polar and N-polar regions as well as the presence of domain walls themselves. The differential phase contrast (DPC)-STEM image (Figure \ref{fig:Npolar}b maps out the deflection of the focused electron beam by crystal defects or magnetic/electric fields. Hence, a distinct color change of about 180$^\circ$ from purple/red to green is observed at the edge of the capacitor indicating the inversion of polar orientation after switching. STEM and HRSTEM investigations (Figure \ref{fig:Npolar}c and \ref{fig:Npolar}d of this edge region show the presence of multiple vertical domain boundaries and in-between regions with superimposed polar structures.

A magnified overview of the partially switched film and its domain structure is presented in the ABF-STEM image of Figure \ref{fig:Npolar}e. A clear change of contrast is observed in an area extending about 10 nm from the Pt electrode interface into the bulk of the film. The boundary of this layer appears fringed, comparable to interfacial inversion domains observed before.\cite{Wolff2025} The local polarity after partial switching is examined by HRSTEM revealing an overall N-polar structure close to the GaN interface and the film bulk (cf. Figure \ref{fig:Npolar}f and \ref{fig:Npolar}g demonstrating ferroelectric polarization inversion). Further, residual M-polar areas are observed at the Pt electrode interface (cf. Figure \ref{fig:Npolar}h). Moreover, the partially switched layer also exhibits vertically extending M-polar regions which have not yet been switched and which are up to 15 nm wide. These domains extend over the whole \alscn film thickness, connecting the GaN interface with the residual M-polar layer at the top interface. The vertical boundaries which separate the M-polar and N-polar regions are characterized by curved tail-to-tail (T-T) domain walls as shown in Figure \ref{fig:Npolar}e and \ref{fig:Npolar}i. 

In comparison to fully switched layers (Figure \ref{fig:Npolar}j), those M-polar interface-bridging regions are not observed and the interfacial M-polar layer is reduced in vertical dimension to about 5\,nm small not-connected M-polar nuclei. In-between the nuclei, areas showing the N-polar motif and areas showing a superposition motif of both polarities are observed (Figure \ref{fig:Npolar}k).\cite{Wolff2024, Kato2024}
Overall these observations imply, in contrast to \alscnM films grown by the MOCVD method, that M-to-N polarization inversion is initiated at the \alscnNospace/GaN interface. This results in a sharp basal N- to M-polar inversion domain boundary at the \alscnNospace/GaN interface, which is not accessible in the case of the MOCVD-grown \alscnMNospace.\cite{Wolff2024, Wolff2025} Another consequence of this is that sputtered \alscn films feature locally tail-to-tail domain walls, while MOCVD-grown \alscnM films only feature head-to-head domain walls. Furthermore, the majority of the polar volume is already switched when applying sub-saturating electric fields. However, a thin layer with M-polar regions is stabilized at the Pt electrode also after applying apparently saturating electric fields (cf. Figure S1).\cite{Gremmel2024}

\subsection{Interfacial switching at the \alscnNospace/GaN interface}

\begin{figure*}[h!]
\centerline{\includegraphics[width=16cm]{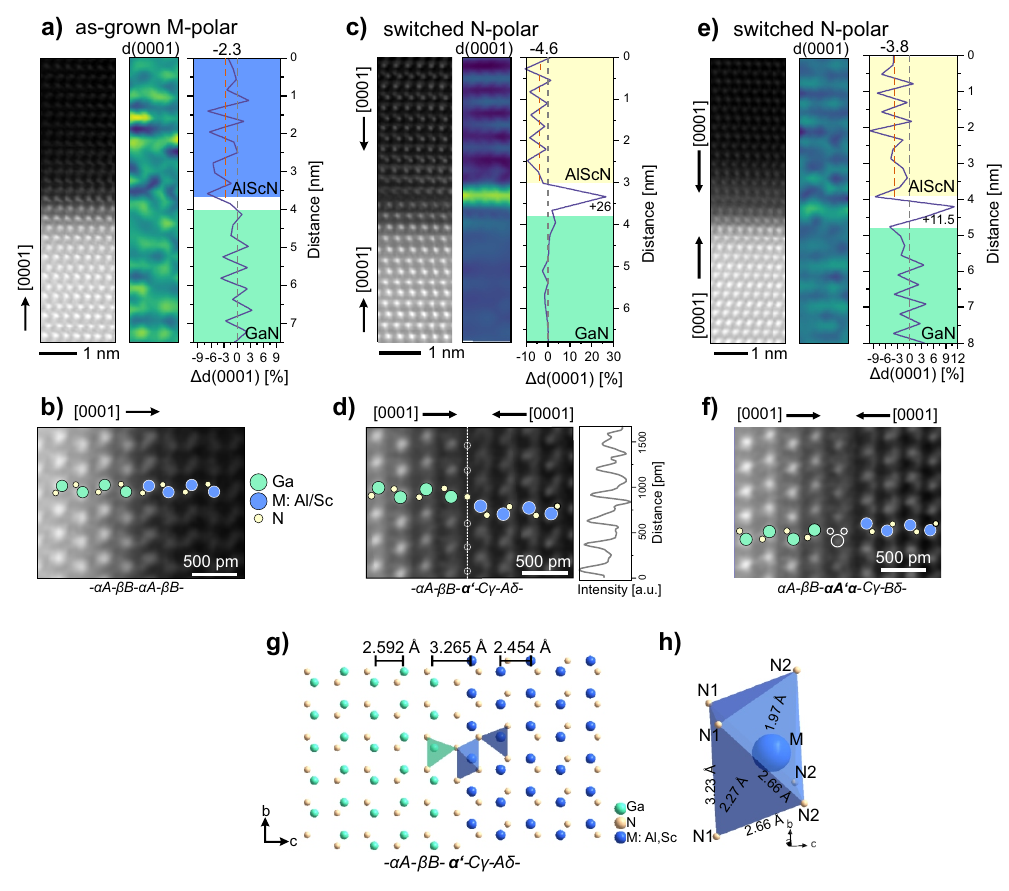}}
\caption{HRSTEM analysis of the atomic interface structures in as-grown M-polar and switched N-polar \alscnNospace/n-GaN heterostructures. a) STEM micrograph, map of the (0001)-metal monolayer separation d(0001) and evolution of relative lattice spacings $\Delta d(0001)$  across the as-grown interface. b) Enlarged HRSTEM image showing the atomic stacking sequence across the as-grown interface. c) STEM micrograph, map of the (0001)-metal monolayer separation d(0001) and evolution of relative lattice spacings $\Delta d(0001)$ across the first configuration of a switched interface. d) Enlarged HRSTEM image showing the atomic stacking sequence across the polarization discontinuity. e) STEM micrograph, map of the (0001)-metal monolayer separation d(0001) and evolution of relative lattice spacings $\Delta d(0001)$ across the second configuration of a switched interface. f) Enlarged HRSTEM image showing the atomic stacking sequence across the polarization discontinuity. 
g) Structure model of the interfacial inversion domain boundary observed in c,d) based on distance measurements between monolayers of Ga and (Al,Sc). Distorted MN6 (M: Al,Sc) octahedra are present at the interface, as shown in h). As 'N1' labeled atomic positions correspond to the anion layer $\alpha'$.\label{fig:interface}}
\end{figure*}

From the HRSTEM micrograph shown in Figure \ref{fig:Npolar}f the formation of a horizontal inversion domain boundary is suggested by the observation of complete polarization inversion close to the GaN interface. Therefore, the atomic structures at the \alscnNospace/GaN interface prior and after ferroelectric switching experiments were examined. HRSTEM micrographs of the interface prior to ferroelectric switching are presented in Figure \ref{fig:interface}a and \ref{fig:interface}c. Using the peak finding method from the Atomap Python library,\cite{Nord2017} the atomic positions of the metal cations (Ga and Al,Sc) was determined from the HAADF-STEM images.
The relative change in interatomic distance $\Delta d(0001)$ at the interface was evaluated by calculating the (0001)-monolayer separation $d(0001)$ while taking the average spacing between the GaN monolayers $d(0001)_{GaN}$ as reference value. 

In Figure \ref{fig:interface}a the HAADF-STEM image of the interface, the corresponding $d(0001)$ map and the difference profile $\Delta d(0001)$ across the interface are shown. The calculated averaged difference within the \alscn monolayers exhibits $\Delta d(0001)$ of  = -2.3\%. We notice that, on average, a -4.4\% smaller \textit{c}-lattice parameter and a -1.2\% in-plane lattice difference was calculated for the film bulk using the RSM and SAED data, respectively. This difference in out-of-plane strain at the interface and the bulk suggests additional interfacial strain from the pseudomorphic coherent growth under slight lattice-mismatch conditions for the present Sc concentration. The observed change of $d(0001)_{AlScN}$ is well correlated with the Z-contrast difference between Ga atomic columns and Al,Sc atomic columns indicating a chemically sharp interface. A magnified view of the atomic structure across the as-grown interface is displayed in Figure \ref{fig:interface}c which shows the repetition of $(\alpha A-\beta B)_n$ structural units for the hexagonal stacking of the M-polar unit cell. In this notation, the Greek letters indicate the anion-(N) atomic positions and the capital letters the cation-(Ga, and Al,Sc) atomic positions.  

After ferroelectric polarization inversion from the M-polar to the N-polar state, the inversion of unit cell polarity in the \alscn layer results in the formation of a horizontal head-to-head (H-H) interfacial polarization discontinuity. 
The image contrast along the interfacial polarization discontinuity is certainly affected by monolayer interface roughness and structural alteration along the horizontal inversion domain boundary and its projection along the film depth.(compare Figure S2, Supporting Information). However, undistorted areas which show an atomically sharp interface across a few to tens of nanometers were identified.
As example, the HRSTEM micrographs presented in Figure \ref{fig:interface}c,d and \ref{fig:interface}e,f show possible atomic configurations of this horizontal interface inversion domain boundary. Again, the monolayer separation $d(0001)$ was calculated from the fitted metal positions revealing a larger relative distance $\Delta d(0001) \sim$ +26\% and $\sim$ +11.5\% between the two metal monolayers at the interface, respectively. In addition, the $\Delta d(0001)$ between the GaN and the N-polar \alscn lattice is determined to -4.6\% and -3.8 \% for both interfaces, which is in agreement with the measured average bulk value and indicates the release of epitaxial lattice strain at the reconstructed interface. This is rationalized by the loss of structural coherence across the interface as displayed in Figure \ref{fig:interface}d showing the alteration of the stacking sequence of $(\alpha A-\beta B)_n$ for the GaN lattice changing into $(C\gamma -A\delta)_n$ for the \alscn lattice with flipped nitrogen positions. Hereby, the atomic structure is changed via a basal plane stacking fault creating a single cubic block with metal layer $A-B-C$ stacking across the interface. The observed increase of the metal to metal layer distance at the interface suggests the presence of an anion-layer $\alpha'$. This assumption is supported by the measured intensity profile within the gap showing evenly spaced peaks of intensity.

The second observed atomic configuration of a monoatomic interface inversion domain boundary is shown in Figure \ref{fig:interface}e,f. Again, a stacking fault along the basal plane is introduced changing the stacking scheme into \textit{A-B-A'-C-B} across the interface, where the A' layer denotes the transition layer between M-polar and N-polar regions. The anion-sub-structure around the A' cation layer is tentatively proposed based on its similarity to the structural superposition motif frequently observed at the polar domain boundaries, i.e., with two projected anion position on each side of the cation.

In Figure \ref{fig:interface}g we propose a model of the atomic structure which is based on the observed interfacial head-to-head boundary and the determined lattice spacings depicted in Figure \ref{fig:interface}c,d . For simplicity, a chemically sharp separation of Ga and (Al,Sc) containing monolayers was assumed. The structural model suggests that the metal positions (layer C) in the first switched N-polar monolayer are coordinated by six nearest neighbors of anions (N1 belong to the anions $\alpha'$ and N2 to $\gamma$, compare \ref{fig:interface}h) forming a distorted octahedron at the interface. These octahedra are connected by common edges in the horizontal direction of the interface and by common corners to the neighboring fourfold coordinated tetrahedra of the wurtzite-type structure. Octahedral coordination of the M(Al,Sc) cation is rarely found in other compounds but exists for instance in the high-pressure phase AlP$_6$O$_{3x}$(NH)$_{3-x}$N$_9$ and of course rock-salt-type \alscnxNospace.\cite{Neudert2017, Satoh2022, Mihalic2023}
 
The formation of planar inversion domain boundaries can be induced by impurity doping of AlN thin films, in which the dopant (e.g., Mg, Sb) cation forms planar ordered monolayers.\cite{Kato2024, Ribic2020} Moreover, it is known that a change of ionic contributions, e.g., via targeted oxygen incorporation into nitrides, can initiate polarization inversion during the growth of AlN-based thin films.\cite{Li2018, Liu2022, Wang2024, Stolyarchuk2018} However, despite a few initial reports, so far no open discussion on interfacial inversion domain boundaries in all-nitride \alscnNospace/GaN heterostructures, either accessed by the growth conditions itself or by \textit{ex situ} ferroelectric switching is available up to date.\cite{Calderon2024, Yazawa2025}

\begin{figure*}[h!]
\centerline{\includegraphics[width=16cm]{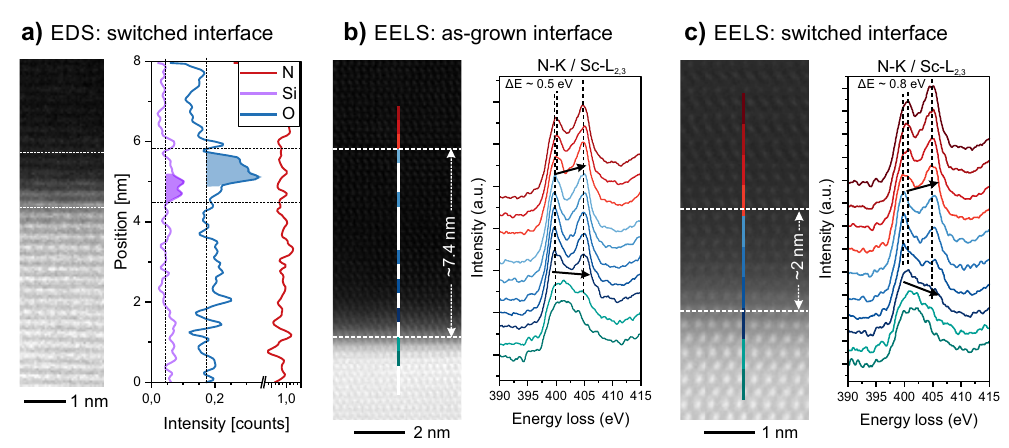}}
\caption{Spectroscopic analyses of the chemical structure at the \alscnNospace/GaN interface. a) STEM-EDS: Si-K, O-K and N-K signal intensity profiles show the presence of Si and O impurity species at the interface. The Si and O intensities are elevated above average, as highlighter by the color-coded areas. b) STEM-EELS: Line-profile measurement of the convoluted N-K | Sc-L$_{2,3}$ edge across the as-grown interface. c) STEM-EELS: Line-profile measurement of the convoluted N-K | Sc-L$_{2,3}$ edge across the switched interface. Arrows indicate the intensity distribution across the two peaks. \label{fig:interface chem}}
\end{figure*}

The chemical structure of the \alscnNospace/GaN interfaces before and after switching was investigated using STEM-EDS mapping and electron energy loss spectroscopy (EELS) presented in Figure \ref{fig:interface chem}. The STEM-EDS intensity profiles (cf. Figure \ref{fig:interface chem}a) and the corresponding elemental maps (cf. Figure S3, Supporting Information) show the elevated signals of oxygen and silicon at the \alscnNospace/GaN interface, but with overall marginal intensity. Silicon is used as n-type dopant to increase the conductivity of the GaN template, whereas the oxygen signal could stem from imperfect substrate cleaning prior to sputter deposition. The total O-K peak intensity at the interface is about twice as large as for the \alscn above the interface. Since the accumulation of oxygen species is known to be influential to the formation of polar inversion domain boundaries it can not be ruled out that it plays a role on the formation of the particular interfaces observed here. 

Further, STEM-EELS investigations of the energetic fine structure of the Sc-L$_{2,3}$ edge across the as-grown and switched interfaces was performed. EELS analysis of \alscnx is naturally facing the difficulty of superimposed N-K (402 eV) and Sc-L$_{2,3}$ (402 and 407 eV) edges as well as a very limited sample space with none yet existing relevant work, which complicates data interpretation of fine structure variations with respect to intensity distributions and energy shifts. 
For \alscnx, the convoluted N-K | Sc-L$_{2,3}$ edge signal shows a characteristic double peak structure which is related to the defined transitions from the 2p$^{3/2}$ and 2p$^{1/2}$ levels into unoccupied 3d-states (L$_{2,3}$) present for the transition metals, so called white-lines.\cite{Graetz2004} In contrast, the N-K edge for GaN only shows one broadened edge profile. For evaluation, the EEL spectra were binned over 3x1 pixels with a total length of 0.6 nm. 
The smoothed spectral data of the N-K | Sc-L$_{2,3}$ edges recorded across the as-grown interface and the polarization discontinuity are presented in Figure \ref{fig:interface chem}b and \ref{fig:interface chem}c, respectively.

As first observation, we note that the edge energy is shifted of about -1 eV  for \alscn with respect to AlN.\cite{Radtke2010}. In case of the as-grown interface, we observe that the characteristic white-line edge profile of Sc-L$_{2,3}$ is evolving from the interface to the bulk. Based on the spectral intensity distribution across the L$_3$ and L$_2$ peaks, two regions can be classified in the majority of our experiments. In the first region, i.e., the interface region, the  L$_3$ peak at lower energy loss exhibits a higher intensity than the L$_2$ peak at higher energy loss; a situation which is reversed in the second region, i.e., the bulk region, showing a higher intensity for the L$_2$ peak. In addition, a small energy shift of about +0.5 eV of the L$_3$ peal is observed with this transition. We noticed, that the onset of the transition of the spectral signature from the interface region to the bulk regions is typically located in the range of $\sim$ 5 - 10 nm from the GaN interface in our experiments.
Further, our EELS investigations across the reconstructed interfaces on switched regions demonstrate that this extended interface region is reduced to about 2 nm as presented in Figure \ref{fig:interface chem}c and Figure S4 (Supporting Information). The locally recorded spectral data show the very similar distribution of peak intensities within the interface region and a clear transition to the bulk region which is accompanied with a slightly larger energy shift of the L$_3$ peak of +0.8 eV.

As mentioned above, we can only speculate on the origin of the observed intensity changes and energy shifts in the convoluted N-K | Sc-L$_{2,3}$ edge profile. One reasonable assumption could be the interrelation with the change of epitaxial lattice strain at the interface, which was discussed with respect to Figure \ref{fig:interface} for both interfaces in detail. A second possibility could arise from the alteration of charge screening effects after switching.

To conclude, the proposed atomic structures and the mechanism of inversion domain boundary formation upon ferroelectric switching require validation by extensive theoretical studies and further observations. Such large interface reconstructions are assumed to be not fully reversible, hence, comparable complex interface structures might be expected after switching back to the M-polar state. Furthermore, implications affecting the magnitude/screening of the polarization discontinuity and so the sheet charge mobility and density of ferroelectric \alscnxNospace/GaN-based interface requires further investigation.

\subsection{Switching pathways in \alscnxNospace/GaN heterostructures produced by MOCVD and sputter epitaxy}

\begin{figure*}[h!]
\centerline{\includegraphics[width=\textwidth]{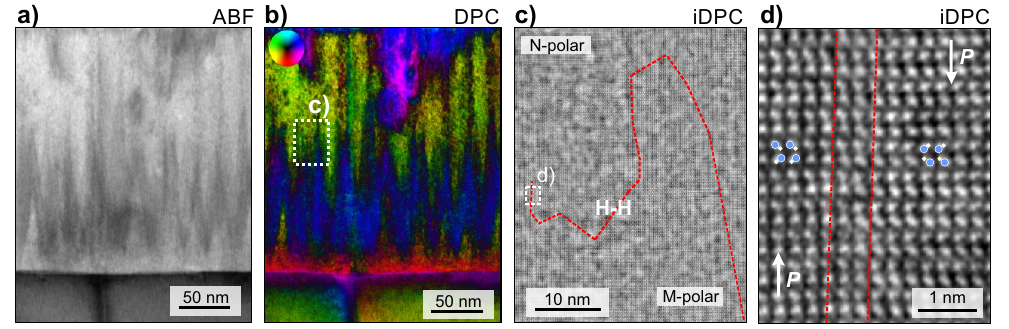}}
\caption{Ferroelectric domain structure in MOCVD-grown \alscnM thin film. a) ABF-STEM image and the corresponding b) DPC-STEM image highlighting the position of ferroelectric domain walls by the center-of-mass shift of the electron beam (blue). The blue color found at the upper part of the film presumably stems from crystalline defects. c) Large-field-of-view iDPC imaging enables to identify the distribution of inclined domain walls between polar domains. The probed region is recorded at one of the tips of the blue spikes, as sketched in b), but is not identical to it. d) HRSTEM image of an inclined domain wall with limited thickness of 1-2 unit cells. The respective region is marked in c).\label{fig:MOCVD}}
\end{figure*}

\begin{figure*}[h!]
\centerline{\includegraphics[width=\textwidth]{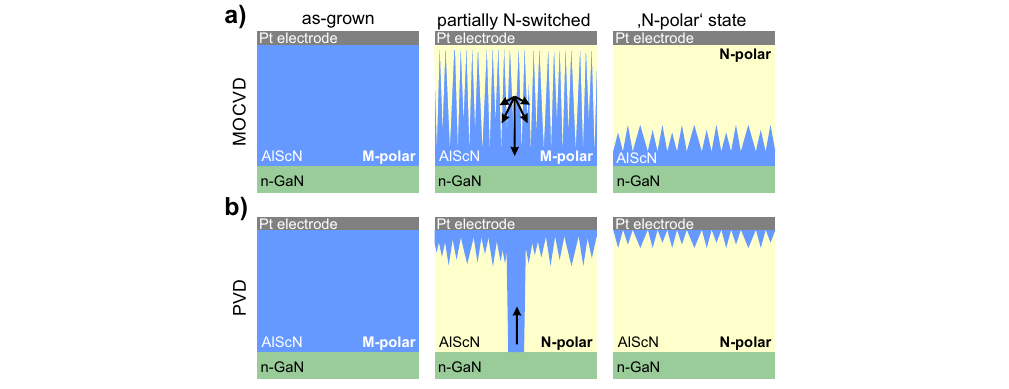}}
\caption{Comparison of the observed switching pathways in a) MOCVD and b) PVD grown epitaxial \alscn thin films on GaN. The black arrows indicate the propagation direction of the N-polar domain front when switching from M-polar to N-polar.\label{fig:mechanisms}}
\end{figure*}

To provide a direct comparison of the established domain patterns in PVD and MOCVD-grown layers we revisit the latter system in Figure \ref{fig:MOCVD} for a short discussion. As outlined in previous works,\cite{Wolff2024, Wolff2025} switching from M- to N-polarity follows via the vertical extension of wedge-shaped domains which propagation is stopped some tens of nanometers above the interface to the GaN, where the MOCVD-grown \alscnM thin film is highly coherent to the M-polar GaN (cf. Figure \ref{fig:mechanisms}a). The ABF-STEM image in Figure \ref{fig:MOCVD}a and the DPC-STEM image of Figure \ref{fig:MOCVD}b show the zig-zag pattern of the inclined domain walls formed in this system. Unlike HAADF-STEM, the integrated (iDPC) signals can further improve image contrast and resolution which allows imaging of atomic structures with light elements and the identification of N- or M-polarity in wurtzite-type structures.\cite{Yücelen2018} 
By recording a large-field-of-view iDPC micrograph with atomic resolution, a good correlation of the zig-zag contrast with the position of domain walls was achieved as sketched onto the micrograph displayed in Figure \ref{fig:MOCVD}c. The dashed line indicates the position of the H-H domain walls separating regions with N-polarity (top) and M-polarity (bottom). A magnified view onto the more vertical domain wall on the left side of the image is shown in Figure \ref{fig:MOCVD}d. The displayed domain wall is approximately one to two unit cells thick.

The domain structure in MOCVD grown \alscnM layers is very different compared to the domain structures observed in the sputtered \alscn layer. The mechanism of polarity inversion is summarized in Figure \ref{fig:mechanisms}b for the sputtered layer.
In contrast to MOCVD-grown layers, the ferroelectric M-to-N polarization inversion in sputtered layers is evidenced to be initiated at the GaN interface, resulting in an atomically-sharp horizontal head-to-head polarization discontinuity at the interface to the M-polar GaN. N-polarity extends through the bulk of the film and stops close to the Pt-electrode interface, resulting in the stabilization of persistent fringed M-polar inversion domain nuclei.\cite{Gremmel2024}

Thinking of a subsequent N-to-M switching event, this observation could resonate with a schematically proposed domain structure model discussed for the ferroelectric ZnMgO.\cite{Yang2024} There, the application of a static electric field to the film surface by a scanning probe microscopy tip lead to the proposed formation of surface-near spike-domains which break through individually to the substrate interface after passing a threshold bias field.
Lately, Yasuoka et al. demonstrated the impact of local lattice strain to initiate the ferroelectric switching by the introduction of additional interfaces and strong compositional gradients between individual layers.\cite{Yasuoka2025} Similarly, monoatomic step edges and non-zero epitaxial strain at the GaN interface together with a more distorted and less coherent crystal structure in sputtered layers could promote polarization flipping at the GaN interface. In contrast, strain-free epitaxial growth of \alscnM achieved by the MOCVD method resulted in an highly coherent crystalline layer at the interface as discussed above, for which the polarity could not be inverted. However, in previous work, we identified vertical chemical defects in the top half of the MOCVD layer, for which small variations of the Sc-concentration were determined. These defects could act as local atomic scale strain centers favoring the observed formation of inversion domains close to the top electrode interface.\cite{Wolff2025} Hence, our direct examinations of ferroelectric domain structures support the observations that polarization switching in \alscnx is strongly interrelated with the nanostructural and chemical properties and its variations within the film.\cite{wang2025, Baksa2024}

The presented understanding of domain structures in sputter deposited ferroelectric \alscnx thin films is highly relevant for the development of future ferroelectric non-volatile memory devices which discriminate between individual polarization states.\cite{Lee2023,Lu2024} Moreover, the revealed pathways of polarization inversion in epitaxial \alscnxNospace/GaN heterostructures provide a starting point for the future development of functional ferroelectric all-III-N device such as FeHEMT.

\section{Conclusions}\label{sec3}

The high crystal quality of nearly lattice-matched \alscn thin films grown by sputter epitaxy on (Si-doped) n-type GaN templates enabled the in-depth examination of field-induced domain structures across a large area using STEM with down to atomic-resolution. The multiscale structural properties were investigated by XRD and TEM and showed a comparable high crystalline quality to lattice-matched \alscnM films grown by the MOCVD method. By studying the polarization domain structures after partial and fully \textit{ex situ} switching events, the switching mechanism could be inferred. It is found that ferroelectric polarization inversion is initiated by the formation of a planar interfacial head-to-head polarization configuration directly at the GaN interface. 
The polarization process is only stopped close to the Pt-electrode interface leaving a thin pinned layer with M-polarity, likely acting as local low energy starting point for subsequent N- to M-polarity inversion. In the partially switched state, the M-polar thin layer appears slightly larger and additional M-polar domains which extend vertically through the whole \alscn volume are evidenced.

The evolution of the atomic and chemical structures upon ferroelectric switching was studied for both the as-grown and switched \alscnNospace/GaN interfaces. Based on the atomic resolution images a structure model of the discontinuity is suggested. Prior to switching, local interfacial epitaxial strain is observed which in concert with the formation of a coherent cubic stacking fault is evidenced to relax to the bulk value after switching. 

Further, we compared the different domain structures and switching mechanisms observed in MOCVD and sputtered \alscnxNospace/GaN heterostructures.
The detailed knowledge on the nanostructure and location of electric-field induced domain boundaries in \alscnx thin films will support the development of future ferroelectric non-volatile memory devices and foster the development of novel ferroelectric III-N heterostructures.

\section{Experimental Details}\label{sec4}

\textbf{Thin film deposition}
For the sputter-deposited heterostructure commercially bought n-type (Si) doped GaN(4 \textmu m)/Sapphire served as template for further film growth. The 100 nm thin \alscn layer was deposited by pulsed-DC co-sputtering from Sc and Al targets at 450 $^\circ$C under N$_2$ and Ar atmosphere by applying 140 W to the Sc target and 860 W to the Al target using an Oerlikon (now Evatec) MSQ 200 multisource system. The 30 nm thin Pt layer was deposited after vacuum break via DC sputtering. Details about the process can be found elsewhere.\cite{Schönweger2022} Electrodes were structured via lithography and ion beam etching (IBE, Oxford Instruments Ionfab 300). Growth details on the ferroelectric MOCVD \alscnM film can be found elsewhere.\cite{Wolff2024}

\textbf{Ferroelectric experiments}
Electrical characterization and pretreatment prior to STEM investigations of individual capacitors was performed by applying a voltage signal with triangular waveform at 1.5 kHz using an aixACCT TF 2000 analyzer. The drive signal was applied to the top electrode. Details on the electrical pre-treatment of the MOCVD-grown \alscnM thin film can be found elsewhere.\cite{Wolff2024}

\textbf{Nanostructure investigations}

\textbf{XRD:}
The X-ray diffraction experiment ($\omega$-scan, RSM) was performed utilizing a Rigaku SmartLab (9 kW, Cu-k$\alpha$). The incident beam was monochromated with a Ge(220)2x monochromator and the diffracted beam was detected with a Hypix-3000 pixel detector. 

\textbf{TEM:}
High-resolution transmission electron microscopy (HRTEM) and high-resolution scanning transmission electron microscopy (HRSTEM) imaging of the sputter deposited \alscnNospace/n-GaN heterostructure was conducted on a probe-corrected transmission electron microscope (NeoARM 200F, Jeol) operated at the acceleration voltage of 200 kV. High-angle annular dark-field (HAADF, 80-220 mrad) and annular bright-field (ABF, 10-20 mrad) STEM detectors were used for resolving the film polarity on the atomic level.
Scan distortions and sample drift during image acquisition were minimized by fast serial recording of multi-frame images followed by post-processing image alignment. The rigid and non-rigid image registration of serial image stacks was performed using the Smart Align \cite{Jones2015} (HREM Research Inc.) plug-in running on the DigitalMicrograph v.3.5.1 (DM) (GatanInc) software. Fourier-filtering of non-rigidly processed STEM micrographs was applied using a simple radiance difference filter (lite version of DM plug-in HREM-Filters Pro/Lite v.4.2.1, HREM Research Inc.) to remove high-frequency noise from the post-processed image. The (0001)-monolayer separation across the interfaces was calculated by the Atomap toolbox using a 2-D Gaussian peak fitting approach for finding atomic column positions.\cite{Nord2017} Electron energy loss spectroscopy (EELS) was performed after tuning the Continuum S spectrometer (Gatan Inc.) to an energy resolution of about 0.55 eV, using an entrance aperture of 5 mm and dispersion of 50 meV/channel. Spectral data was recorded using the dual-EELS acquisition mode and serial acquisition along line profiles, thereby distribution the dose across several nanometers in width. Data analysis was done on the DM software. For plotting, the spectral data was smoothed using the Savitzky-Golay approach.

Integrated differential phase contrast (iDPC) HRSTEM and EDS experiments were conducted on the double-corrected transmission electron microscope (Themis Z, Thermo Fisher) operated at the acceleration voltage of 300 kV and equipped with the Super-X EDS system. The electron beam deflection across domain walls was evaluated by calculating the center-of-mass (COM) shift of the scattered intensity via segmented detectors with collection angles between 6 -23 mrad. The convergence angle of the electron beam was 30 mrad. DPC mapping to detect the position of domain walls was performed using the electron beam with a convergence angle of 10 mrad and the collection angle range is 3 to 17 mrad.

\newpage

\newpage
\subsection*{Acknowledgments}
This collaborative work was enabled through funding by the Federal Ministry of Education and Research (BMBF) under project no. 03VP10842 (VIP+ FeelScreen) and in project ProMat\_KMU “PuSH” Grant Number 03XP0387B and the Deutsche Forschungsgemeinschaft (DFG, German Research Foundation) 
Project-ID 286471992 - SFB 1261 as well as Project-ID 458372836 and Project-ID 448667535.
Funded by the European Union (FIXIT, GA 101135398). Views and opinions expressed are however those of the author(s) only and do not necessarily reflect those of the European Union or the European Research Council Executive Agency. We acknowledge the Karlsruhe Nano Micro Facility (KNMFi) at KIT for providing access to DPC-STEM based capabilities.
The authors acknowledge Dr. Stefano Leone and Dr. Isabel Streicher for the development and MOCVD growth of ferroelectric AlScN thin films.





\bibliography{wileyNJD-AMA}



\end{document}